# Auger recombination in narrow band quantum well $Cd_xHg_{1-x}Te/Cd_yHg_{1-y}Te$ heterostructures

V.Ya.Aleshkin[1,2], G. Alymov[3], A.V.Antonov[1,2], A.A.Dubinov[1,2], V.V.Rumyantsev[1,2], and S.V.Morozov[1,2]

[1]Institute for Physics of Microstructures of RAS, 603950 Nizhny Novgorod, Russia

[2]Lobachevsky University of Nizhny Novgorod, 603950 Nizhny Novgorod, Russia

[3]Laboratory of 2d Materials for Optoelectronics, Moscow Institute of Physics and Technology, Dolgoprudny 141700, Russia

**Abstract**

We present detailed theoretical and experimental studies of Auger recombination in narrow-gap mercury cadmium telluride quantum wells (HgCdTe QWs). We calculate the Auger recombination probabilities as functions of non-equilibrium carrier density, temperature and composition of quantum wells taking into account the complex band dispersions and wave functions of the structures. Our theory is validated by comparison with measured kinetics of photoconductivity relaxation in QW with band gap of 76 meV at a temperature of 77 K. We find good agreement of theory and experiment using a single fitting parameter: the initial density of non-equilibrium carriers. The model is further used to optimize the composition of QWs and find the most suitable conditions for far-infrared lasing. Particularly, for band gap of 40 meV (lasing wavelength λ=31 µm) the lasing is favored in QWs with 6.5% cadmium fraction. We also find that at very large non-equilibrium carrier densities, the main recombination channel is associated with emission of two-dimensional plasmons and not with Auger process.

## I. Introduction

Auger recombination is one of the main obstacles in the development of long-wavelength interband semiconductor lasers. In recent years, outstanding results were achieved in the development of far-infrared laser structures based on CdHgTe quantum wells (QWs), particularly, stimulated emission up to 20 µm was observed under optical pumping [1]. The possibility of such emission can be attributed to highly symmetric Dirac-like spectrum of electrons and holes in CdHgTe QWs that raises the threshold of non-radiative Auger recombination [2]. However, to date there are only a few studies considering the Auger recombination in HgTe QWs [2–4]. Refs. [3,4] dealt with quantum wells with band gaps exceeding 200 meV. In such QWs, the Auger process differs significantly from that in narrow-gap structures with a band gap below 50 meV. In addition, the Auger recombination rate in Refs. [2, 3] was calculated only at the threshold electron-hole density at which stimulated emission occurs. Thus, the studies of Auger recombination in narrow-gap CdHgTe quantum wells are lacking completeness.

The most practical problem that did not yet receive due attention is the optimal width and composition of HgTe QWs and surrounding barriers delivering favorable conditions for stimulated emission. This problem is particularly important for the 30-40 µm wavelength range where the operation of GaAs-based quantum cascade lasers is impossible due to phonon absorption [5].

The present work is devoted to the theoretical study of Auger recombination processes in narrow-gap quantum well $Cd_xHg_{1-x}Te/Cd_yHg_{1-y}Te$ heterostructures. We calculate the dependences of the recombination probability on the momentum of charge carriers and their concentration. We compare our results with the experimentally observed time dependences of photoconductivity under conditions when the predominant recombination mechanism is Auger process. We search for the optimal parameters of QWs for lasing at a photon energy of 40 meV. Taking the $Cd_{0.065}Hg_{0.935}Te$ QW as example ($E_G$ = 40 meV), we show that recombination with emission of two-dimensional plasmons becomes the main recombination channel at the nonequilibrium electron densities above $1.4\times10^{11}$ cm$^{-2}$.

The paper is organized as follows. Section II presents the computational model for Auger recombination in CdHgTe QWs. Section III is devoted to the validation of our model by its comparison with experimental data on photoconductivity relaxation. In section IV, the parameters of CdHgTe QWs are optimized for laser action in the far-infrared range. Section V briefly addresses another type of recombination due to Coulomb interaction of carriers: the recombination with emission of two-dimensional plasmons. The main conclusions are summarized in Section VI.

## II. Computational model for the Auger recombination probability

The spectrum and states of electrons in CdHgTe QWs were calculated using the four-band Kane model with inclusion of strain effects. To simplify the calculations, the effects of symmetry lowering at the heterointerfaces and the absence of inversion symmetry were neglected. All calculations were performed for structures grown on the [013] crystallographic plane, since most experimental studies [1,6] are carried out on such structures. The explicit form of the Kane

Hamiltonian for the case under consideration as well as the method for finding the electronic states can be found in [7]. Wave functions of electron states have the form $\Psi(\mathbf{k},s,\mathbf{r}) = \exp(i\mathbf{k}\boldsymbol{\rho})\psi_s(\mathbf{k},z)/\sqrt{S}$, where $\boldsymbol{\rho}$ is the radius-vector lying in the quantum well plane, $\mathbf{k}$ is the 2d wave vector, and the z-axis is points along the normal to the quantum well plane.

Fig. 1 shows the calculated spectrum of the quantum well for the structure with $Cd_{0.08}Hg_{0.92}Te$ quantum wells of 7.8 nm width and $Cd_{0.65}Hg_{0.35}Te$ barriers; the same structure was used in experimental study of non-equilibrium carrier kinetics. The effective band gap at 77 K equals 76 meV. Under excitation conditions, nonequilibrium carriers fill only the main quantization subbands. The final state of the final 'hot' electron in the Auger process always belongs to the lower electron subband [2]. This feature is due to the fact that the band gap in such structures is less than the distance between the size quantized subbands in the conduction band (see Fig. 1). In addition, thresholdless Auger processes involving two holes and one electron are absent at the temperatures under consideration [8]. These features of Auger process greatly facilitate the calculation of its probability and allow us to restrict the consideration to recombination involving two electrons and one hole [2] (CCCH process). The probability of the CCCH process for an electron with momentum $\mathbf{k}$ and quantum number $s$ (which labels the two-dimensional subband and the spin indices) can be written as:

$$W(\mathbf{k},s) = \frac{2\pi}{\hbar} \sum_{\mathbf{k}_1,\mathbf{k}'} \sum_{s_1,s',s_f} \frac{S^2}{(2\pi)^4} \iint d^2k_1 d^2k' \left|V(\mathbf{k},s,\mathbf{k}_1,s_1,\mathbf{k}',s',\mathbf{k}_f,s_f)\right|^2 \times$$
$$\times f(\mathbf{k}_1,s_1)[1-f(\mathbf{k}',s')][1-f(\mathbf{k}_f,s_f)] \times \delta(\varepsilon_c(\mathbf{k}) + \varepsilon_c(\mathbf{k}_1) - \varepsilon_c(\mathbf{k}') - \varepsilon_v(\mathbf{k}_f)) \quad (1)$$

where $\hbar$ is the Planck's constant, $\mathbf{k}_1, s_1\ \mathbf{k}', s'$ are the wave vectors and quantum numbers of the second electron before and after recombination, respectively, $\mathbf{k}_f, s_f$ are the wave vector and quantum number of the final state belonging to the valence band, $\varepsilon_j(\mathbf{k})$ is the electron dispersion law in the $j$-th band, $f(\mathbf{k},s)$ is the Fermi-Dirac electron distribution function, $V$ is the matrix element of electron-electron interaction:

$$V(\mathbf{k},s,\mathbf{k}_1,s_1,\mathbf{k}',s',\mathbf{k}_f,s_1') = \int d\mathbf{r} d\mathbf{r}_1 \Psi^+(\mathbf{k}',s',\mathbf{r})\Psi^+(\mathbf{k}_f,s_1',\mathbf{r}_1)U(\mathbf{r},\mathbf{r}_1)\Psi(\mathbf{k},s,\mathbf{r})\Psi(\mathbf{k}_1,s_1,\mathbf{r}_1) -$$
$$- \int d\mathbf{r} d\mathbf{r}_1 \Psi^+(\mathbf{k}',s',\mathbf{r})\Psi^+(\mathbf{k}_f,s_1',\mathbf{r}_1)U(\mathbf{r},\mathbf{r}_1)\Psi(\mathbf{k},s,\mathbf{r}_1)\Psi(\mathbf{k}_1,s_1,\mathbf{r}) \quad (2)$$

$U$ is the potential of the screened Coulomb interaction of electrons. The first and second terms in (2) account for direct and exchange electron-electron interactions. Note that in Auger process under consideration, the momentum conservation can be written as $\mathbf{k} + \mathbf{k}_1 = \mathbf{k}' + \mathbf{k}_f$.

To simplify the calculations, we will neglect the difference between the dielectric constants of the quantum well and barriers. In addition, we will assume that the characteristic impact parameter in the interaction of electrons is much larger than QW width. Using these approximations and neglecting quantum effects in screening, we obtain the following expression for the matrix element of direct

$$V_d(\mathbf{k},s,\mathbf{k}_1,s_1,\mathbf{k}',s',\mathbf{k}_f,s_1') = \frac{2\pi e^2}{S\kappa} \int dz dz_1 \frac{\exp(-|\mathbf{k}-\mathbf{k}'||z-z_1|)}{(|\mathbf{k}-\mathbf{k}'|+1/\lambda)} \times$$
$$\psi_{s'}^+(\mathbf{k}',z)\psi_{s_f}^+(\mathbf{k}_f,z_1)\psi_s(\mathbf{k},z)\psi_{s_1}(\mathbf{k}_1,z_1)$$
(3)

and exchange interactions:

$$V_{ex}(\mathbf{k},s,\mathbf{k}_1,s_1,\mathbf{k}',s',\mathbf{k}_f,s_1') = \frac{2\pi e^2}{S\kappa} \int dz dz_1 \frac{\exp(-|\mathbf{k}-\mathbf{k}'||z-z_1|)}{(|\mathbf{k}-\mathbf{k}_f|+1/\lambda)} \times$$
$$\psi_{s'}^+(\mathbf{k}',z)\psi_{s_f}^+(\mathbf{k}_f,z_1)\psi_s(\mathbf{k},z_1)\psi_{s_1}(\mathbf{k}_1,z).$$
(4)

In the above expressions, $e$ is the electron charge, $S$ is the area of the quantum well, $\kappa$ is the dielectric constant, and $\lambda$ is the characteristic screening length given by:

$$1/\lambda = \frac{2\pi e^2 a}{\kappa} \tag{5}$$

If electrons and holes occupy only the lowest conduction and upper valence quantization subbands, respectively, the expression for a has the form:

$$a = \frac{2}{(2\pi)^2 k_B T} \int d^2k \left\{ \frac{\exp\left(\frac{\varepsilon_c(k)-F_c}{k_B T}\right)}{\left[1+\exp\left(\frac{\varepsilon_c(k)-F_c}{k_B T}\right)\right]^2} + \frac{\exp\left(\frac{F_v+\varepsilon_v(k)}{k_B T}\right)}{\left[1+\exp\left(\frac{F_v+\varepsilon_v(k)}{k_B T}\right)\right]^2} \right\}, \tag{6}$$

where $T$ is the temperature, $k_B$ is the Boltzmann constant, and $F_j$ is the chemical potential in the $j$-th band.

Known the Auger recombination probability, the net recombination rate $R_{Auger}$ can be found from:

$$R_{Auger} = \frac{1}{2}\sum_{k,s} W(k,s) f(k,s) \tag{7}$$

The coefficient ½ in (7) takes into account the double inclusion of the same characteristics of electrons when summing in (7) and (1).

### III. Comparison with experiment

In order to check the adequacy of our model for Auger recombination, we compare the theoretically calculated and experimentally observed kinetics of photoconductivity. The experiment was carried out at $T = 77$ K. The structure was grown at the Institute of Semiconductor Physics SB RAS in the group of N.N. Mikhailov and S.D. Dvoretsky on the [013] plane of GaAs substrate with ZnTe (50 nm) and CdTe (10 μm) buffer layers. The technology of structure growth is described in [9]. The structure contained 5 $Cd_{0.08}Hg_{0.92}Te$ quantum wells of 7.8 nm width separated by $Cd_{0.65}Hg_{0.35}Te$ barriers.

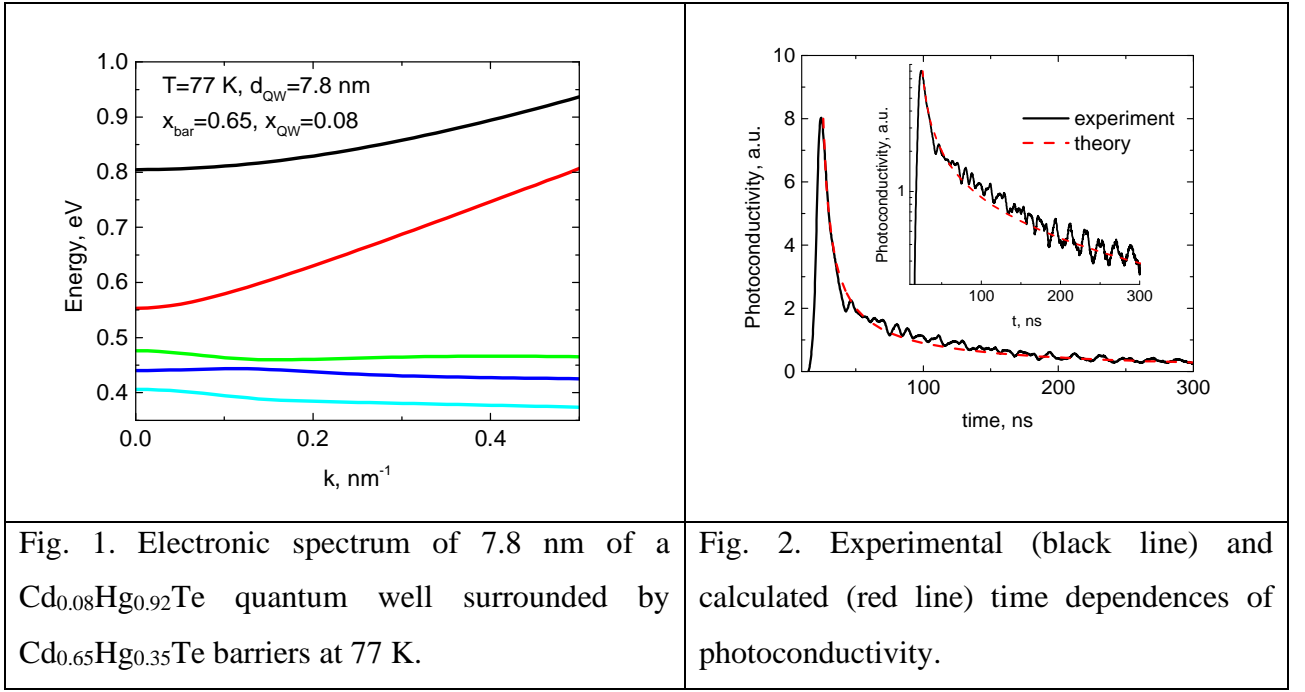

Fig. 1. Electronic spectrum of 7.8 nm of a $Cd_{0.08}Hg_{0.92}Te$ quantum well surrounded by $Cd_{0.65}Hg_{0.35}Te$ barriers at 77 K.

Fig. 2. Experimental (black line) and calculated (red line) time dependences of photoconductivity.

The experimentally study of nonequilibrium carrier concentration decay is based on the relaxation of the interband photoconductivity (PC) signal. The sample was excited by a Solar parametric light generator (Minsk, Belarus). The signal was recorded on a Le Croy digital oscilloscope with an upper cutoff frequency of 1 GHz and amplified by an amplifier with a bandwidth of 400 MHz. The time resolution of the method was determined by the pulse duration being equal to 7 ns. We used the radiation with 9.5 μm wavelength and a pulse energy of up to ~ 10 μJ in a spot of 7 mm diameter. The measurement scheme is described in detail in [10-12]. The observed time dependence of photoconductivity for a structure with a 7.8 nm quantum well is shown in Fig. 2 with black solid line.

Our model of photoconductivity is based on the assumption of direct proportionality between conductivity and the density of excited carriers $n$. The latter is obtained by direct numerical integration of balance equation:

$$\frac{dn}{dt} = -R_{Auger} - R_{rad} \qquad (8)$$

where $R_{rad}$ is the rate of radiative recombination calculated using the method described in [13]. To highlight the similarity of calculated and measured dependences at long times, we show both on a logarithmic scale in the inset to Fig. 2.

Since the initial density of nonequilibrium carriers $n(0)$ is unknown in the experiment, it was used as a fitting parameter in the calculations. The best agreement of the experimental and calculated traces is observed for $n(0) = 8.5 \times 10^{10} cm^{-2}$. The figure shows good agreement between the measured and calculated dependences up to about 300 ns. At longer times, the photoconductivity could not be measured with reasonable accuracy. During the observation time, the photoconductivity dropped by about 25 times.

It is possible to show that recombination kinetics in the QW under consideration is dominated by Auger and not by radiative recombination processes. To prove this fact, we show the density dependences of the Auger and radiative recombination probabilities averaged over the distribution of electrons in Fig. 3. Using the initial concentration of nonequilibrium carriers $n(0)$ and comparing Figs. 2 and 3, we conclude that Auger recombination is dominant in almost the entire interval of observation. Only at the final stages of photoconductivity relaxation, the probability of radiative recombination becomes comparable with that of Auger recombination. Good agreement between simulated and experimentally measured relaxation kinetics allow us to conclude that our model is adequate for describing Auger recombination in CdHgTe QWs with ~70 meV band gap.

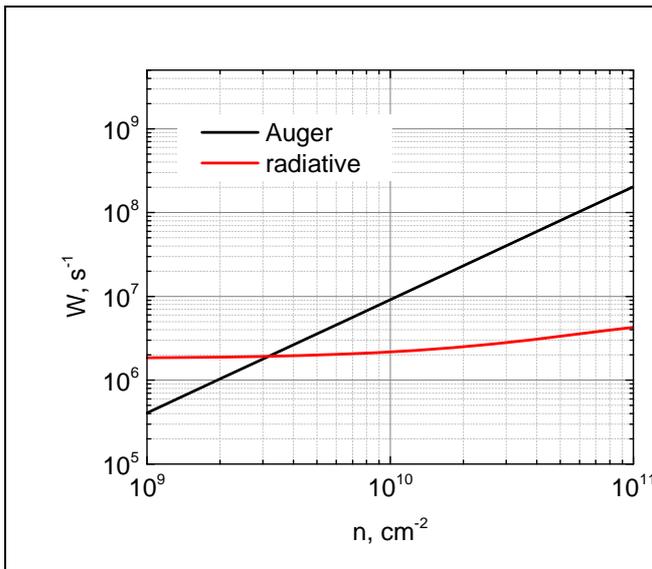

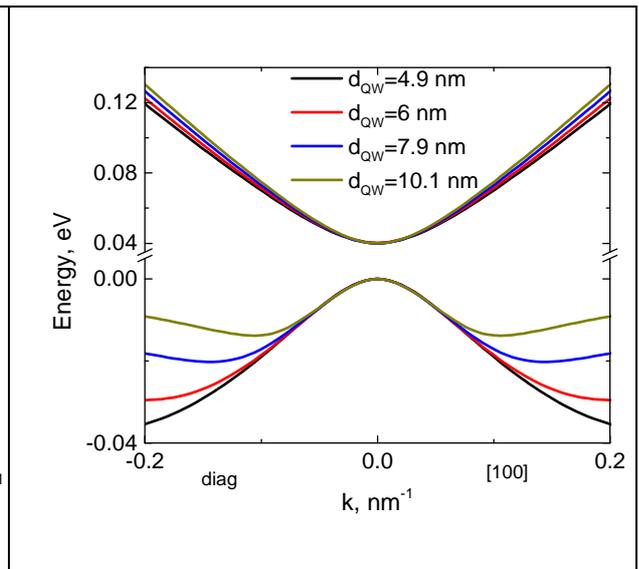

Fig. 3. Dependences of the probability of recombination for the Auger process (black line) and radiative transition (red line) on nonequilibrium carrier concentration.

Fig. 4. Electronic spectra of four structures at 8K. The direction of the wave vector, denoted by diag, corresponds to the diagonal between the directions [100] and [03-1].

### IV. Optimization of quantum well parameters for stimulated emission

The question of the optimal QW parameters for laser action in the far-infrared range is of great applied importance. To answer this question, we fix the band gap to 40 meV (31 µm wavelength) and study the effect of QW composition on Auger and radiative recombination. The composition of the barriers is fixed to $Cd_{0.7}Hg_{0.3}Te$. As shown earlier [8], such a barrier composition is optimal from the point of view of the maximum threshold Auger recombination energy. The probabilities of radiative and Auger recombination were calculated for four quantum wells with cadmium fractions of 0%, 3%, 6.5%, and 9% and thicknesses of 4.9 nm, 6 nm, 7.9 nm, and 10.1 nm, respectively. The calculation was performed for 8 K temperature at which stimulated emission from a 6.5% structure was observed [1].

Figure 4 shows the electronic spectra of these structures. The figure shows clearly that an increase in the cadmium fraction in the quantum well leads to a slight decrease in the effective electron mass in the conduction band. The effective hole mass at the Γ- point of the Brillouin zone remains almost unchanged thereby. Another effect of cadmium admixture is the 'growth' of the side extrema in the valence band.

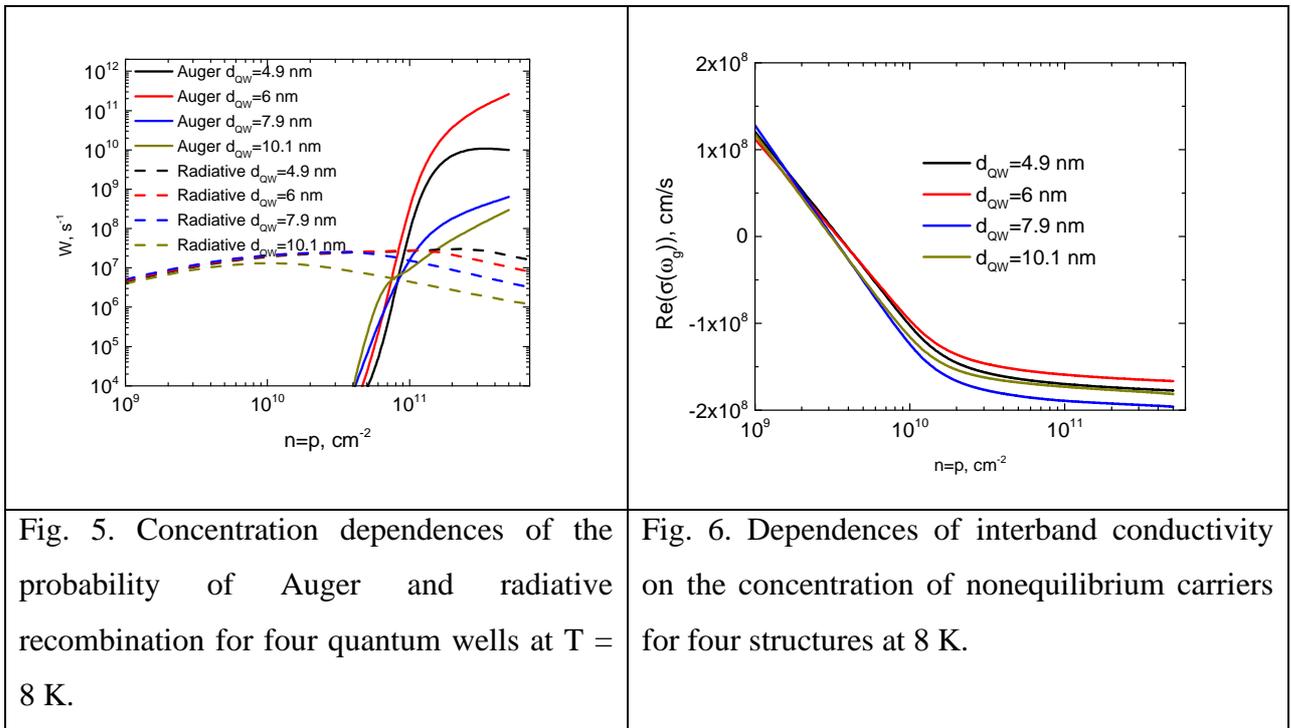

| Fig. 5. Concentration dependences of the probability of Auger and radiative recombination for four quantum wells at T = 8 K. | Fig. 6. Dependences of interband conductivity on the concentration of nonequilibrium carriers for four structures at 8 K. |
|---|---|

The calculated density dependences of the radiative and Auger recombination probabilities are shown in Fig. 5. As seen from the figure, the dependence of Auger recombination probability differs significantly from $n^2$ expected for Maxwellian electron-hole statistics. The reasons for the deviation from the $n^2$-dependence are the degeneracy of electron-hole system and the density dependence of the screening length.

Remarkably, the calculated density dependence of Auger recombination probability in the structure with 4.9 nm QW is non-monotonous and drops down at n> $2 \times 10^{11}$ cm$^{-2}$. This fact can be attributed to the enhancement of screening at large densities. The density dependences of radiative recombination probability are always non-monotonous, but the reasons for decrease are different. They lie in the filling of side extrema in the valence band [13] with holes that cannot participate in radiative processes due to absence of electrons with correspondingly large momenta.

For laser action purposes, the most important carrier density $n^*$ corresponds to almost equal probabilities of Auger and radiative recombination (intersection of solid and dashed lines of the same color in Fig. 5). As this density is exceeded, Auger recombination prevails and leads to the heating of the electron gas. This, in turn, increases the Auger recombination rate and decreases the radiative recombination rate. Therefore, the optimal conditions of stimulated emission are achieved in the structures with **highest** $n^*$. Fig. 5 shows that for the structures under consideration, the maximum value of $n^*$ is achieved in a quantum well with cadmium fraction of 6.5% and width of 7.9 nm.

Another important figure of merit for lasing applications of QWs is the value of the real part of the interband conductivity at given density of nonequilibrium carriers. This quantity determines the laser gain [14] for radiation with frequency $\omega$ and is given by

$$\mathrm{Re}\,\sigma(\omega) = \frac{e^2}{4\omega\pi} \sum_{s,s'} \int d^2k \, |\mathbf{v}(\mathbf{k},s;\mathbf{k},s')_{cv} \boldsymbol{\eta}|^2 \, \delta(\varepsilon_c(\mathbf{k}) - \varepsilon_v(\mathbf{k}) - \hbar\omega)[f_v(\mathbf{k}) - f_c(\mathbf{k})] \qquad (9)$$

where $\mathbf{v}(\mathbf{k},s;\mathbf{k},s')_{cv}$ is the interband matrix element of the velocity operator, $\boldsymbol{\eta}$ is the unit vector along the direction of the electric field of the electromagnetic TE wave.

The dependences of $\mathrm{Re}\,\sigma(\omega_g)$ on nonequilibrium carrier density for the structure with $\hbar\omega_g = E_g = 40$ meV are shown in Fig. 6 (averaged over the direction $\boldsymbol{\eta}$ in the plane of the quantum well). It can be seen from the figure that the amplification of radiation ($\mathrm{Re}\,\sigma(\omega_g) < 0$) sets in at carrier density of about $3\times10^9$ cm$^{-2}$, at which the interband population inversion is realized. The figure also shows that the highest absolute value of the gain in a wide range of densities is realized in the structure with a cadmium fraction of 6.5% in the quantum well. The latter fact is associated with the large value of the interband matrix velocity elements in this structure.

All above calculations allow us to conclude that the most favorable conditions for stimulated emission are realized in the structure with a cadmium fraction of 6.5% in the quantum well.

The dependences of the Auger recombination probability on the electron wave vector in the 'optimal' QW are shown in Fig. 7 for five concentrations of nonequilibrium carriers. It can be seen from the figure that an increase in the wave vector of an electron and an increase in concentration increase the probability of Auger recombination. "Roughness" of the lines in Fig. 7 is due to the inaccuracy of calculating the integrals in (1) over wave vectors by the Monte Carlo method.

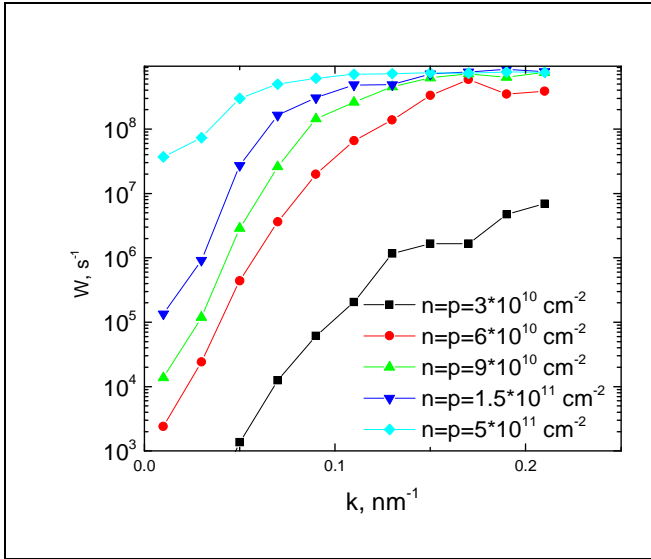 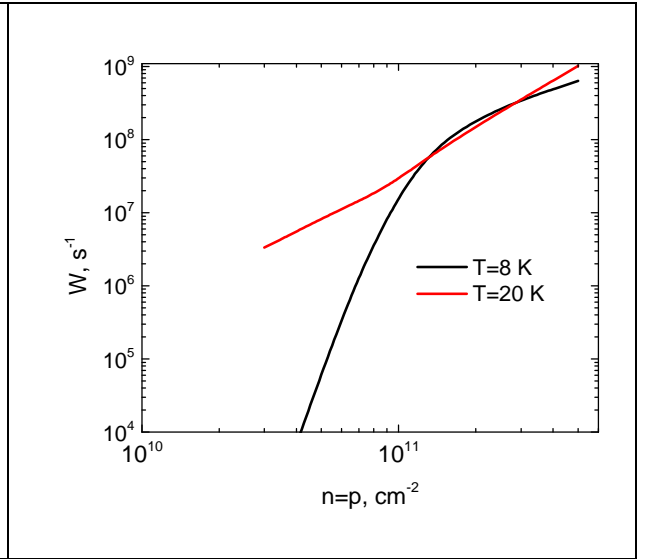

Fig. 7. Dependence of the probability of Auger recombination on the wave vector of an electron for five concentrations of minority carriers in a quantum well with a cadmium fraction of 6.5% at 8 K.

Fig. 8. Dependences of the probability of Auger recombination on the concentration of nonequilibrium carriers in a quantum well with 6.5% cadmium at 8 K and 20 K.

To conclude this section, we discuss the variation of the Auger recombination probability with increasing temperature. Fig. 8 shows the dependences of this probability on the nonequilibrium carrier density in an optimal QW at 8 K and 20 K. The density dependence of $W$ is close to $n^2$ at 20 K. In general, increase in temperature leads to an increase in the probability, apart from the interval $1.3 \times 10^{11}$ cm$^{-2}$ < n < $2.8 \times 10^{11}$ cm$^{-2}$.

## V. Plasmon-assisted recombination

At sufficiently high concentrations of nonequilibrium carriers, recombination with the emission of two-dimensional plasmons can start dominating in narrow-gap quantum wells [15]. In this section, we illustrate this statement by the example of a quantum well 7.9 nm wide with 6.5% cadmium fraction. To calculate the recombination rate due to emission of two-dimensional plasmons, we use the approach developed in [15]. The spectrum of two-dimensional plasmons in this approach is calculated with account of the spatial dispersion of the electron gas susceptibility. Recombination with the emission of plasmons is activated above a certain threshold concentration of nonequilibrium carriers [16]. Below the threshold, this type of recombination is impossible due to energy-momentum conservation constraints. For the structure under consideration, the threshold concentration is approximately equal to $1.4 \times 10^{11}$ cm$^{-2}$.

The density dependences of the radiative, Auger, and plasmon-assisted recombination probabilities are shown in Fig. 9. The figure illustrates that at above a certain critical concentration, recombination with the emission of two-dimensional plasmons becomes the dominant recombination mechanism. In the structure under consideration, the effective refractive index for two-dimensional plasmons exceeds 100. Therefore, the reflection coefficient of a plasmon from the edge of the structure is close to unity. If no special precautions are taken for effective plasmon 'removal' from the structure, the energy released upon recombination will be spent on heating the electron and hole gases. This circumstance will prevent the realization of the inverted population of the bands and the stimulated generation of plasmons.

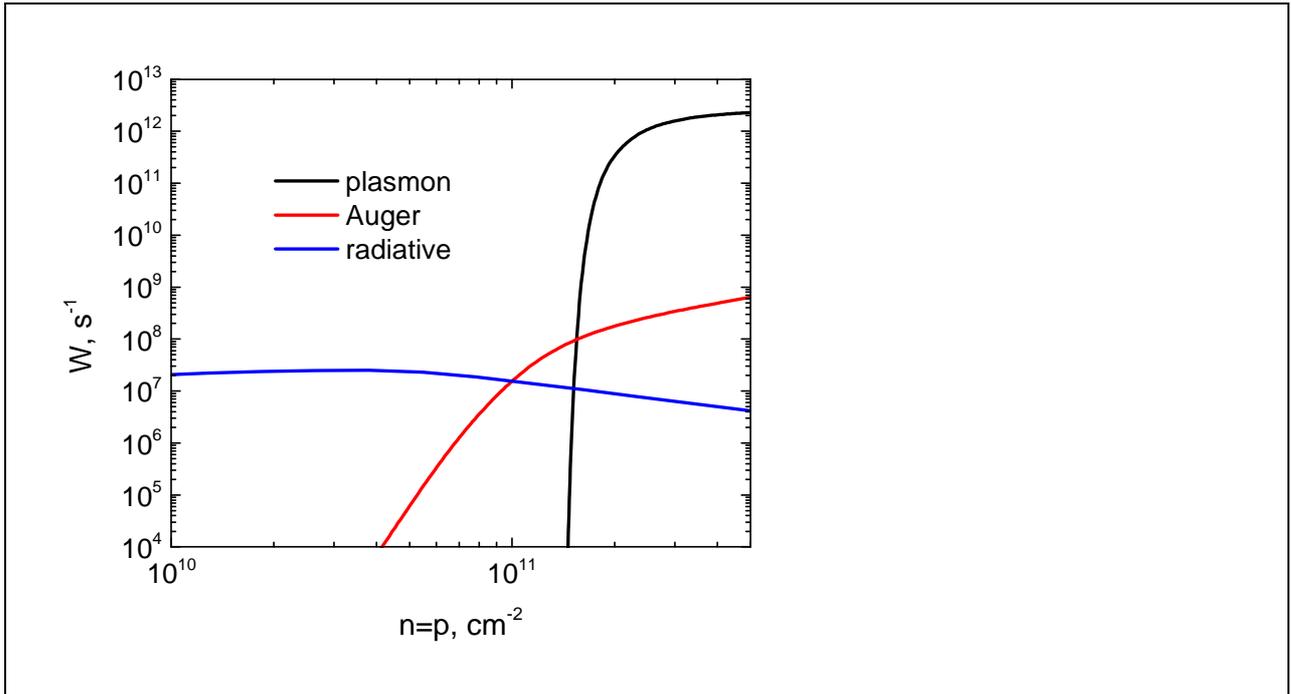

Fig. 9. Dependences of the probabilities of recombination on the concentration of nonequilibrium carriers for three recombination processes in a quantum well with a cadmium fraction of 6.5% at 8 K.

### VI. Conclusion

In conclusion, we briefly list the main results of the work.

1. A model was developed for calculating the probability of Auger recombination in narrow-gap quantum well $Cd_xHg_{1-x}Te/Cd_yHg_{1-y}Te$ heterostructures.

2. A comparison was made of the experimentally observed and calculated kinetics of photoconductivity under conditions when the main mechanism of recombination is Auger recombination. Good agreement between theory and experiment was obtained.

3. A search for the optimal parameters of quantum wells for generating stimulated radiation with a quantum energy of 40 meV at T = 8 K has been carried out. It is shown that the optimal fraction of cadmium in the quantum well is 6.5%.

4. It is shown that in a quantum well with a cadmium fraction of 6.5% when the concentration of nonequilibrium carriers exceeds $1.4 \times 10^{11}$ cm$^{-2}$, the main mechanism of recombination is recombination with the emission of two-dimensional plasmons.

**Acknowledgments**

Part of the work on the calculation of Auger recombination was supported by the Russian Foundation for Basic Research (project 18-02-00362) and the Ministry of Science and Higher Education of the Russian Federation (grant 075-15-2020-797 (13.1902.21.0024)). Part of the work on the calculation of plasmon recombination was supported Russian Science Foundation (project 20-42-09039). The authors thank D. Svintsov for proofreading of the manuscript.